# MATHEMATICAL DISCOVERY OF NATURAL LAWS IN BIOMEDICAL SCIENCES: A NEW METHODOLOGY

**Leonid Hanin**

921 S. 8th Avenue, Stop 8085, Department of Mathematics and Statistics, Idaho State University, Pocatello, ID 83209-8085, USA

E-mail: hanin@isu.edu

**Abstract.** As biomedical sciences discover new layers of complexity in the mechanisms of life and disease, mathematical models trying to catch up with these developments become mathematically intractable. As a result, in the grand scheme of things, mathematical models have so far played an auxiliary role in biomedical sciences. We propose a new methodology allowing mathematical modeling to give, in certain cases, definitive answers to systemic biomedical questions that elude empirical resolution. Our methodology is based on two ideas: (1) employing mathematical models that are firmly rooted in established biomedical knowledge yet so general that they can account for any, or at least many, biological mechanisms, both known and unknown; (2) finding model parameters whose likelihood-maximizing values are independent of observations (existence of such parameters implies that the model *must not* meet regularity conditions required for the consistency of maximum likelihood estimator). These universal parameter values may reveal general patterns (that we call *natural laws*) in biomedical processes. We illustrate this approach with the discovery of a clinically important natural law governing cancer metastasis. Specifically, we found that under minimal, and fairly realistic, mathematical and biomedical assumptions the likelihood-maximizing scenario of metastatic cancer progression in an individual patient is invariably the same: *Complete suppression of metastatic growth before primary tumor resection followed by an abrupt growth acceleration after surgery*. This scenario is widely observed in clinical practice and supported by a wealth of experimental studies on animals and clinical case reports published over the last 110 years. The above most likely scenario does not preclude other possibilities e.g. metastases may start aggressive growth before primary tumor resection or remain dormant after surgery.

## 1. Introduction: A New Methodology of Mathematical Modeling

Physics has universal laws and by the grace of Nature they have precise mathematical form [1]. In molecular biology, various microscopic processes that are essentially physical or chemical, such as the action of molecular motors or kinetics of biochemical reactions, can also be described by fairly precise mathematical models. By contrast, in the systems biology of cells, organs and tissues as well as in medicine, no universal quantitative laws have ever been discovered. In their stead, biomedical sciences deal with qualitative or quantitative statements about biological objects or processes that are based on measurements, observations, theoretical arguments, rejection of alternatives, experience or intuition and are believed to be true for most biological organisms or patients of certain kind. We will call such statements, perhaps for lack of better term, *natural laws*. As one example, consider a statement, often made by medical doctors, that systematic consumption of certain foods, e.g. read meat products, increases the risk of a specific cancer. Two related factors make verification of this statement difficult. First, biological mechanisms behind this putative cause and effect relationship remain largely unknown (although hypotheses abound). Second, cancer risk-specific effects of diet are most likely individual, so that the above statement may be true for some people and false for others. Typically, statements like this are approached empirically based on the comparison of the number of people who have developed cancer among those with and without a particular diet habit or by comparing pre-diagnosis diets of cancer patients to those in the general population. Such comparisons, however, are methodologically problematic because cancer risk is influenced by many other factors correlated with diet such as genetic make-up, family history of cancer, age, BMI, amount of exercise, exposure to environmental or occupational hazards, and many others. The question we take up in this article is: *Can natural laws in biomedical sciences be formulated in precise mathematical terms and discovered or confirmed unequivocally through mathematical modeling and statistical analysis?*

Mathematical modeling of biomedical systems, such as individual cells, multicellular organisms, patients or their populations, faces two principal challenges. The first is enormous complexity of biomedical processes. As biology and medicine uncover layer after layer of molecular mechanisms of life and disease, mechanistic mathematical models attempting to catch up with these developments gradually lose their mathematical tractability and eventually computational feasibility. The second challenge is great variability and heterogeneity of biological systems. This typically requires mathematical models to have stochastic components and include parameters accounting for individual variation.

We propose here an alternative methodology of mathematical modeling in biomedical sciences. The new methodology is based on two ideas. The first is to build mathematical models that, on the one hand, are firmly grounded in the well-established biomedical knowledge but, one the other hand, eschew mechanistic details and are general enough to account for *any,* or at least many, mechanisms, both known and unknown. Simplicity and generality of such models, representing two of the mathematics' most notable hallmarks and sources of strength, offer a greater chance to enable closed form solutions and rigorous theoretical analysis than complex and highly specific mechanistic models.

Mathematical models in biology and medicine depend on numerical and functional parameters, and the former can be estimated from observations. One of the most appealing ways to do so is to select parameter values that maximize the model-based probability of observations or, in the case of continuous observables, their joint probability density function (pdf). This time-tested method of parameter estimation, known as the Principle of Maximum Likelihood, goes back to Gauss, Lagrange, Euler, Daniel Bernoulli and Laplace [2].

In general, the likelihood-maximizing values of parameters depend on observations. However, it may happen that some of them are *universal*, that is, the same for all admissible data sets. These universal parameter values are therefore a property of the model rather than a function of observations. Thus, if the model is general enough and depends only on a minimal set of reasonable assumptions, then one can hope to have thus discovered a general pattern in biomedical phenomena, i.e. a natural law. This likelihood-maximizing state of the biomedical system or pathway of the process does not exclude other possibilities; however, they are less likely to occur. This is the second idea underlying our new approach to mathematical modeling of biomedical phenomena. To the best of our knowledge, the new methodology has never been employed in the past.

Before we proceed, one statistical remark is in order. If data is *generated by the model* with a fixed set of parameters then one would expect the maximum likelihood estimators (MLEs) of numerical parameters based on this data to converge, as the sample size increases indefinitely, to the parameter values used to generate the data, which would prevent the existence of universal MLEs. However, this convergence property, called *consistency of the MLE*, is only true if the model satisfies certain regularity conditions, see e.g. [3, p. 202]. Therefore, allow data-independent MLEs, the model *must not* meet the regularity conditions.

As an illustration of the new methodology, we apply it to cancer metastasis and obtain a definitive answer to a question of great clinical importance. Specifically, application of the Principle of Maximum Likelihood to an extremely general individual-patient model of cancer progression, see Section 3, has led us to a surprising discovery that the most likely scenario under the model is invariably the same: C*omplete suppression of the growth of metastases while primary tumor is in place followed by an abrupt acceleration of their growth following primary tumor resection*, see [4] for more details. Importantly, this phenomenon was discovered in animals models more than a century ago [5-7], later confirmed in many other experimental studies on animals, and supported by countless clinical observations, see comprehensive reviews [8-12]. This finding also represents a common knowledge among veterinarians.

## 2. Primary Tumor and Metastases: A Brief Overview

Over the last few decades, cancer has become a global pandemic. As one example, if the current trend in cancer incidence and mortality continues unabated more than a quarter of the world population and about 39% of people in the US (40% of men and 38% of women) will develop cancer over their lifetime and about half of them will die of cancer. The most clinically significant aspect of solid cancers is metastasis, which accounts for about 90% of all cancer-related deaths. In the grand scheme of things, mathematics and statistics have played so far an auxiliary role in cancer biology and oncology. As we will show below, they can give definitive answers to clinically important questions whose empirical resolution by biomedical sciences has been heretofore elusive.

As every practicing oncologist knows, cancer patients seldom present with clinically manifest metastases. For example, even for renal cell carcinoma, an insidious kidney cancer that frequently progresses asymptomatically until advanced stages of the disease, 25-35% represents a historical maximum for the fraction of patients having clinical metastases at diagnosis (nowadays, due to widespread diagnostic imaging, this fraction is about 10-15%). Yet in a large fraction of cancer patients, metastases surface within weeks, months or years after surgical removal of the primary tumor. *Is this just a coincidence or is there a cause and effect relationship between (a) the presence of the primary tumor and absence of clinical metastases; and (b) primary tumor resection and surfacing of metastases?* Although numerous experiments on animals accumulated over more than 110 years provide ample empirical evidence supporting the suppressing effect of the primary tumor and accelerating effect of its resection on metastasis, direct confirmation of this striking phenomenon and understanding of its causes in humans have evaded biomedical scientists so far, for three reasons. Firstly, subclinical micrometastases and critical microevents that trigger their origination and progression to detectable secondary tumors are largely *unobservable*. Secondly, such randomly occurring formative microevents are separated by months, years or even decades from their clinical manifestations. Finally, experimental verification of the above tentative natural law is restricted by ethical considerations. This makes the use of mathematical models indispensable for establishing qualitative and quantitative relationships between primary tumor and metastases.

Over a few recent decades, cancer biology and oncology underwent a revolutionary transformation brought about by the discoveries of (1) mutational origins of cancer and genomic instability of cancer cells; (2) their metabolic plasticity; (3) aberrant regulation of cell cycle, proliferation and apoptosis in cancer cells; (4) the pivotal role of angiogenesis in tumor progression; (5) the significance of primary and secondary tumor dormancy; (6) complex interactions of cancer cells with the immune and endocrine systems; (7) a critical role of tumor microenvironment; (8) an important role of inflammation in cancer origination and progression; (9) identification of stem-like cancer cells; and (10) the importance of selection processes that bring about heterogeneity of cancer cell populations and frequently lead to failure of cancer treatment. Although they have all considerably changed the landscape of cancer research, they have not, unfortunately, brought about a revolution in cancer treatment, for a simple reason that many other important aspects of cancer still remain unknown. That is why when answering systemic questions about cancer like the one posed above one cannot rely on mathematical models based on specific biological mechanisms – many of them have not yet been discovered! This, combined with insurmountable complexity of any mathematical model accounting for all the known mechanisms of cancer origination and progression, calls for a different approach; the one outlined in the Introduction serves precisely this purpose.

The study of the interaction between various tumors within the same host has a long history starting with a pioneering work of 1906 by Paul Ehrlich [5] who discovered experimentally that tumor-bearing animals frequently reject second tumors with which they were inoculated or suppress their growth. This and many other later studies have proved beyond reasonable doubt that large tumors may inhibit the growth of smaller ones, see comprehensive reviews [8, 11]. In particular, it was observed in animal experiments that primary tumor suppresses the growth of metastases [6-8, 11]. It comes as no surprise, then, that primary tumor resection is followed in many cases by rapid outgrowth of metastases. This effect was first reported in [6, 7] and later confirmed in many experimental studies and supported by numerous clinical case reports, as reviewed in [8-12]. Additionally and importantly, any surgical intervention has metastasis-promoting effects of its own due to a variety of reasons including (1) increased local and systemic production of growth and angiogenesis factors required for wound healing; (2) transient inflammation that acts as universal cancer promoter; and (3) temporary immunosuppression that may cause metastases to escape immune surveillance.

Solid primary tumors originate from a single malignant clonogenic cell. Unlike normal cells of the same tissue origin, cancer cells can evade internal, local and systemic signals that regulate cell cycle and apoptosis. Tumor growth is enabled by supply of nutrients, oxygen and growth factors initially delivered through diffusion. However, according to a fundamental discovery made by Judah Folkman in the 1970s [13], under such conditions a tumor can only reach a microscopic diameter of about 1-2 mm. To enable its further growth, the tumor has to induce angiogenesis, i.e. grow a capillary network that will provide a more efficient supply of essential chemicals. If the effects of angiogenesis inhibitors are stronger than those of angiogenesis promoters then the tumor will be in a state of dormancy characterized by a dynamic equilibrium between cell proliferation, death and quiescence. By contrast, if due to internal or external factors the balance shifts in favor of angiogenesis the tumor will vascularize and start an aggressive growth. This event, viewed below as instantaneous for the purpose of mathematical modeling, will be called primary tumor *inception.*

A primary tumor may start shedding bloodborne metastases as soon as it becomes vascularized, i.e. following inception. Because the minimum diameter for clinical detection of tumors is about 5 mm, the aforementioned Folkman's limit for the diameters of avascular tumors suggests that at the time of inception a primary or secondary tumor may still be undetectable.

Metastasis is a complex multi-stage process [14]. A metastasis-generating cell has to separate itself from the primary tumor, degrade extracellular matrix, migrate to a blood vessel, intravasate, survive through the period of free circulation, extravasate, and invade a host site. A solitary metastatic cell lodged in a secondary site may remain dormant for months, years or even decades [15]. Under favorable conditions, it may start to grow, induce angiogenesis and eventually enter the clinical stage. Thus, metastasis is a highly inefficient and selective process in that only a tiny fraction of cancer cells shed off the primary tumor give rise to clinically detectable secondary tumors. Just as for primary tumor, the start of active growth of a vascular metastasis will be termed *inception*.

### 3. An Individual-Patient Mathematical Model of Metastatic Cancer

An extremely general mathematical model described below focuses on well-established clinically important processes associated with metastasis while avoiding explicit description of their causes and mechanisms. For previously proposed particular cases of the model as well as its extensions and parametric versions, see [4, 16-21]. The model combines a deterministic law of growth of metastases in a given secondary site (typically lungs, liver, bones, brain or soft tissues) with a stochastic description of the processes that clearly depend on chance [22], such as shedding of metastases off the primary tumor, their survival, and dormancy in the given site. The "trick" here is that the rate of metastasis shedding and the distribution of metastasis latency times are essentially arbitrary.

**Primary tumor inception and observables.** Let T be the unobservable age of primary tumor inception. Its significance is that it represents the earliest time when primary tumor may start shedding metastases. Denote by V the age at primary tumor resection and by W a later age at which several metastases were detected and measured in a given secondary site. Thus, $0 < T < V < W$, see Fig. 1. Let m be the smallest measurable tumor size (i.e. the number of cells) in the site in question. Denote by $X_1, X_2,..., X_n$, where $m < X_1 \leq X_2 \leq ... \leq X_n$, the sizes of the detected metastases. Such data can be obtained through laborious reading of CT/PET images, as exemplified in [16-21].

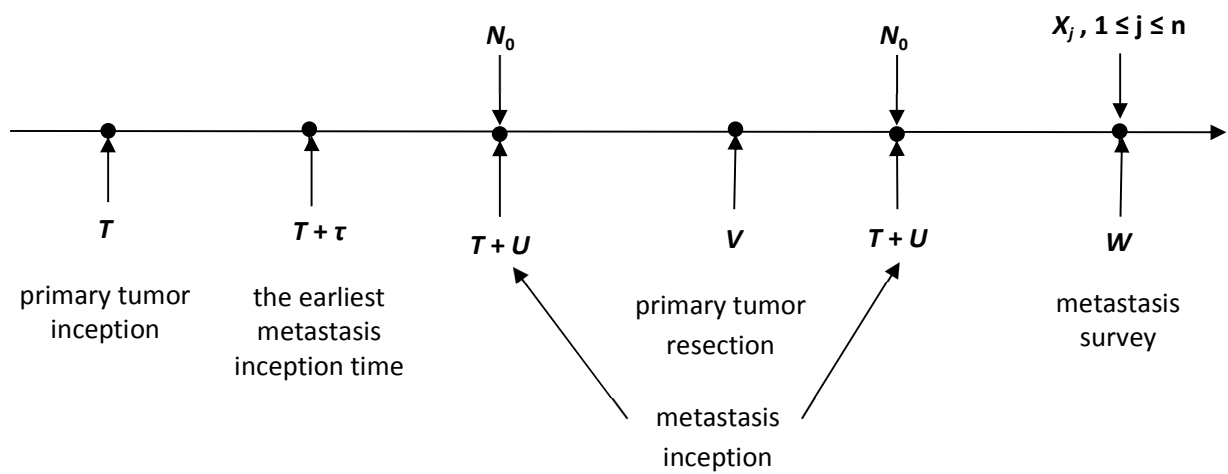

**Fig. 1**

Fig. 1. Timeline of the natural history of metastatic cancer and its treatment in an individual patient. All times are represented as ages counted from the birth of the patient. T+U is the age at inception of a particular metastasis. Note that the inception may occur either before or after surgery. Numbers with arrows above the timeline represent sizes of the metastasis at the indicated time points.

**Metastasis generation.** It is assumed that metastasis shedding off the primary tumor is governed by a Poisson process with rate r(t), where time t is counted from the age T. Recall that this means, by definition, that the numbers of metastases shed over non-overlapping time intervals are stochastically independent and that the number of metastases shed over any interval [s, t] is Poisson distributed with the expected value $\int_s^t r(u)du$, see e.g. [23, pp. 270-273]. We assume that function r is continuous and positive on [0, V-T) and that r(t) = 0 for t > V-T. Implicit in the latter condition are the assumptions that the resected primary tumor is non-recurrent and that secondary metastasizing (that is, formation of "metastasis of metastasis") to a given site, both from other sites and from within, is negligible, which is commonly believed to be the case [24].

**Progression of metastases.** Metastases shed by the primary tumor are assumed to survive and give rise to clinically detectable secondary tumors in a given site independently of each other with the same probability q > 0; we will call such metastases *viable*. Therefore, see e.g. [25, pp. 257-259], inception of viable metastases in the site of interest is governed by the Poisson process with rate μ(t) = qr(t). Furthermore, each viable metastasis is assumed to spend some random time between detachment from the primary tumor and inception in the secondary site, termed *metastasis latency time*. This time includes dormancy of a metastasis as a solitary cancer cell and as an avascular micrometastasis lodged in the site of interest. Latency times of viable metastases bound for a given site are thought of as independent and identically distributed (iid) random variables with some continuous pdf f and the corresponding cumulative distribution function F. Then, the resulting delayed Poisson process is again a Poisson process with rate $\int_0^t \mu(s)f(t-s)ds$, see e.g. [26]. Finally, we assume for simplicity that the growth of all vascular metastases in the site in question starts from the same initial size, $N_0$, at inception. This implies that the duration of metastatic latency is bounded below by a certain minimal time τ > 0 during which a metastasis can reach the size $N_0$ if it starts to grow from a single cell at a maximum possible rate immediately after seeding. Accordingly, we assume that f(t) = 0 for 0 ≤ t ≤ τ and f(t) > 0 for τ < t ≤ W–T. We think of τ as a sufficiently small number so that T+τ < V, see Fig. 1. Recall that a typical threshold m corresponds to metastatic diameter of about 5 mm while the diameter of metastasis at inception is close to the Folkman's avascular limit of about 1-2 mm. Therefore, m > $N_0$.

**Growth of metastases.** To answer our question about the effects that primary tumor and its removal exert on the rate of growth of metastases, we have to parameterize one component of our model: the laws of growth of metastases before and after primary tumor resection. Quite often, clinical metastases are detected early when they are likely to be still in the exponential phase of their growth. Therefore, we assume that after inception all viable metastases in the given site grow exponentially with two net growth rates, $\gamma_0$ and $\gamma_1$, not necessarily distinct, before and after primary tumor resection, respectively. Denote by U the inception time of a particular metastasis relative to the age, T, of primary tumor inception, see Fig. 1, and let $\Psi_U(t)$ be the size of the metastasis at time t counted from age T+ U. Recall that $\Psi_U(0) = N_0$. The exponential growth function g(t) = N exp{γt}, t ≥ 0, has the following self-similarity property: $g(t) = g(t_0)\exp\{\gamma(t-t_0)\}$, $t \geq t_0$, for every $t_0 > 0$. This allows us to combine the two exponential growth laws to obtain

$$\Psi_U(t) = \begin{cases} N_0\exp\{\gamma_0 t\} & \text{if } T+U < V \text{ and } 0 \leq t \leq V-T-U \\ N_0\exp\{\gamma_0(V-T-U)\}\exp\{\gamma_1[t-(V-T-U)]\} & \text{if } T+U < V \text{ and } t > V-T-U \\ N_0\exp\{\gamma_1 t\} & \text{if } T+U \geq V \text{ and } t \geq 0 \end{cases} \quad (1)$$

**Effects of surgery.** Surgery is assumed to affect metastases only through the rate of their growth (and not through a change in the distribution of their latency times). This assumption is plausible because dormant tumors are commonly viewed as refractory to various treatments including surgery, radiation and chemotherapy [27].

## 4. Distribution of the Sizes of Metastases

Let X be the size of a particular metastasis with inception time U detected in a given site and measured at age W. Then according to (1)

$$X = \Psi_U(W-T-U) = \begin{cases} N_0\exp\{\gamma_0(V-T-U)\}\exp\{\gamma_1(W-V)\} & \text{if } T+U < V \\ N_0\exp\{\gamma_1(W-T-U)\} & \text{if } T+U \geq V \end{cases}$$

where W–T–U is the total metastasis *progression time* from inception to detection. An important observation is that the function that relates X to the progression time is independent of U:

$$X = \Psi(W-T-U) \quad (2)$$

where

$$\Psi(y) = \begin{cases} N_0\exp\{\gamma_1 y\} & \text{if } 0 \leq y \leq W-V \\ \exp\{\gamma_0[y-(W-V)]\}\exp\{\gamma_1(W-V)\} & \text{if } W-V < y \leq W-T-\tau \end{cases} \quad (3)$$

Note that function $\Psi$ is increasing, continuous, piecewise differentiable, and satisfies the condition $\Psi(0) = N_0$. The inverse function, $\delta = \Psi^{-1}$, is given by

$$\delta(x) = \begin{cases} \dfrac{1}{\gamma_1}\ln\dfrac{x}{N_0} & \text{if } N_0 \leq x \leq A \\ \dfrac{1}{\gamma_0}\ln\dfrac{x}{N_0} + (W-V)\left(1-\dfrac{\gamma_1}{\gamma_0}\right) & \text{if } A < x \leq M \end{cases} \quad (4)$$

where $A = N_0\exp\{\gamma_1(W-V)\}$ is the size at age W of a hypothetical metastasis whose inception occurred at the time of surgery and $M = N_0 \exp\{\gamma_0(V-T-\tau)\}\exp\{\gamma_1(W-V)\}$ is the theoretical upper limit for the size of a metastasis measured at age W, i.e. the size of a hypothetical metastasis that was shed off the primary tumor at time T and whose inception with size $N_0$ occurred over the shortest latency period possible, i.e. at age $T+\tau$, see Fig. 1. Note that $N_0 < A < M$.

According to formula (2) the sizes of metastases at age W are a fixed non-random transformation $\Psi$ of their inception times which, under our model, follow a Poisson process. Then, applying to the case at hand a well-known theorem about the conditional joint distribution of the event occurrence times in a Poisson process conditional on their number, see e.g. [23, pp. 264–265], we obtain the following result.

**Theorem 1.** *The sizes $X_1 \leq X_2 \leq \ldots \leq X_n$ of metastases in a given site that are detectable at age W are equidistributed, given their number n, with the vector of order statistics for a random sample of size n drawn from the distribution with pdf*

$$p(x) = \delta'(x) \frac{\int_0^{\min\{W-T-\delta(x), V-T\}} r(s) f[W-T-\delta(x)-s]\,ds}{\int_0^{\min\{W-T-\delta(m), V-T\}} r(s) F[W-T-\delta(m)-s]\,ds}, \quad m < x < M, \quad (5)$$

*and $p(x) = 0$ for $x \notin (m, M)$, where $\delta(x)$ is given by (4).*

For a proof of Theorem 1 in the case $r(t) = \alpha t^\theta$, $0 \leq t < V-T$, with some constants $\alpha > 0$ and $\theta \geq 0$, see [14]; the proof in the general case is obtained by a minor modification. Distribution (5) depends on functional parameters r, f and numerical parameters $\gamma_0, \gamma_1$. Notice, however, that it is free of metastasis survival probability q and sample size n. The patient- and/or site-specific biological parameters T, $N_0$, $\tau$, m are assumed fixed.

In view of (3) we have $W-T-\delta(x) \leq V-T$ if and only if $x \geq \Psi(W-V) = N_0 \exp\{\gamma_1(W-V)\} = A$. Therefore, in the case $A \leq m$ we have

$$p(x) = \frac{1}{\gamma_0 x} \frac{\int_0^{W-T-\delta(x)} r(s) f[W-T-\delta(x)-s]\,ds}{\int_0^{W-T-\delta(m)} r(s) F[W-T-\delta(m)-s]\,ds}, \quad m < x < M, \quad (6)$$

while if $A > m$ then

$$p(x) = \begin{cases} \dfrac{1}{\gamma_1 x} \dfrac{\int_0^{V-T} r(s) f[W-T-\delta(x)-s] ds}{\int_0^{V-T} r(s) F[W-T-\delta(m)-s] ds}, & m < x < A \\[2em] \dfrac{1}{\gamma_0 x} \dfrac{\int_0^{W-T-\delta(x)} r(s) f[W-T-\delta(x)-s] ds}{\int_0^{V-T} r(s) F[W-T-\delta(m)-s] ds}, & A < x < M \end{cases} \quad (7)$$

The first and second formulas in (7) account for metastases whose inception occurred after and before primary tumor resection, respectively. From (4) we find that $\delta(M) = W-T-\tau$. This implies, in view of our assumption about the support of pdf f, that $p(M-) = 0$. Thus, pdf p is continuous at point M. By contrast, pdf has jump discontinuity at point m as well as at point $A > m$, unless $\gamma_0 = \gamma_1$. That is why model (5) does not meet regularity conditions for the consistency of MLEs. Finally, because $W-T-\delta(m) > W-T-\delta(M) = \tau$, the denominator in formulas (6) and (7) is positive.

Because the Poisson process of metastasis inception is non-stationary, the site-specific sizes of metastases at age W do not form a random sample from any probability distribution. However, it follows from Theorem 1 that the distribution of any symmetric (i.e. rearrangement-invariant) statistic based on the observations $X_1, X_2,..., X_n$ is identical to the distribution of the same statistic based on a random sample of size n drawn from the pdf p given by formula (5). One such symmetric statistic is the joint likelihood of observations

$$L(X_1, X_2,..., X_n) = \prod_{k=1}^{n} p(X_k) \quad (8)$$

Thus, pdf p serves, for the purpose of likelihood computation, as a surrogate of the non-existent distribution of the "size of a detectable metastasis at age W." It is also worth mentioning that in the case $A > m$ if one of the observations $X_k$ coincides with A then, in order to maximize the likelihood, the larger of the two one-sided limits of pdf p at $x = A$ should be employed.

## 5. The Main Result

Because cell cycle duration has a lower limit, the growth rate of any cell population is bounded above. Therefore, the net growth rates $\gamma_0$ and $\gamma_1$ are bounded above by some constant C. However, due to cell death and quiescence they may approach zero. Let $G = (0, C) \times (0, C)$ be the set of admissible pairs $(\gamma_0, \gamma_1)$. Denote by

$$B = \frac{\ln \dfrac{X_n}{N_0}}{W - V}$$

the rate of growth of a hypothetical metastasis whose inception occurred at age V and that reached size $X_n$, the largest among the observed sizes of metastases, at age W. The following theorem specifies the likelihood-maximizing parameters $\gamma_0, \gamma_1$. For a detailed proof, see [4].

**Theorem 2.** *For any fixed functional parameters r, f and for every data set $X_1, X_2,..., X_n$ such that $m < X_1 \leq X_2 \leq ... \leq X_n$, the supremum of the likelihood function $L(\gamma_0, \gamma_1)$ given by (8) over set G is infinite. Furthermore, if $L(\gamma_0^{(n)}, \gamma_1^{(n)}) \to \infty$ for some sequence $(\gamma_0^{(n)}, \gamma_1^{(n)})$ in G then $\gamma_0^{(n)} \to 0$ and $\gamma_1^{(n)} \to B$.*

Theorem 2 implies that the likelihood-maximizing estimate of parameter vector $(\gamma_0, \gamma_1)$ is (0, B). Notice that the estimate $\gamma_0 = 0$ is independent of the data! By contrast, the estimate $\gamma_1 = B$ is a function of the largest observation $X_n$. Thus, according to the model, the most likely metastasis progression scenario for a patient with n metastases detected at age W > V in a given site is as follows: *All those (say, k) metastases, $1 \leq k \leq n$, whose inception occurred before resection of the primary tumor remained essentially dormant ($\gamma_0 \to 0$) before surgery while after surgery they were growing at a rate $\gamma_1$ close to B and by age W reached sizes close to $X_n$. Additionally, the remaining n - k metastases whose inception occurred after surgery were growing at about the same rate $\gamma_1 = B$ and by age W reached more widely spread smaller sizes $X_1, X_2, ..., X_{n-k}$.* The model predicts, therefore, that with sufficiently high probability the observed data set will contain, along with widely spread sizes of metastases whose inception occurred after surgery, some number $k \geq 1$ of more tightly clustered larger sizes of metastases whose inception occurred prior to surgery. Interestingly, this somewhat unexpected pattern was actually observed in all 19 patients the author had analyzed including three breast cancer patients [17], twelve prostate cancer patients [18], and four renal cancer patients [21].

## 6. Discussion

Universality of our MLE findings that $\gamma_0 = 0$ and $\gamma_1 > 0$ implies that *the most likely pathway of metastatic cancer progression in an individual patient involves complete suppression of metastatic growth while primary tumor is in place followed by its abrupt acceleration following primary tumor resection*. This scenario is consistent with millions of clinical cases, represents a common knowledge among veterinarians, and may be explained by a number of biological mechanisms [4, 11, 12, 28, 29]. This does not preclude other scenarios; however, they are expected to occur with smaller probability. In particular, some metastases may escape the suppressing effects of the primary tumor before surgery while others may remain dormant long after surgery. Although the model that has led us to these conclusions is extremely general, its modest assumptions are still only an approximation to reality. As a result, the qualitative natural law formulated above is expected to only manifest as a general trend in a patient- and cancer type-specific manner.

Derivation of the above clinically significant natural law from a very general mathematical model may create an impression of a rabbit extracted from a hat. Such an impression, however, does not account for the fact that the model is grounded in three most significant aspects of metastatic cancer discovered in the second half of the 20[th] century. The first is highly selective, multi-stage, stochastic nature of metastasis seeding [14, 22]. The second is that dissemination of bloodborne metastases may occur long before primary tumor becomes clinically detectable. This striking finding was discovered, initially in the case of breast cancer, through a lifetime of extensive experimental work, clinical observations and numerous clinical trials by Bernard Fisher and his colleagues [30, 31]. The third is the significance of metastatic dormancy in the form of solitary cancer cells or avascular micrometastases [15, 32, 33]. What we have shown mathematically is that under very mild assumptions of our model the above natural law holds true for every metastatic progression process involving (1) Poisson dissemination interrupted by resection of the primary tumor; (2) random independent survival of metastases; (3) their latency with random iid durations; and (4) their exponential growth with two fixed rates, one before and one after removal of the primary tumor. Thus, the natural law is rooted in very basic temporal and kinetic aspects of metastasis.

The validity and generality of the discovered natural law depends on how well our model assumptions agree with reality. These assumptions fall into two categories: mathematical and biomedical. Mathematical premises include, first, homogeneity and independence assumptions, i.e. that metastases bound for a given site survive independently of each other with the same probability, that their latency times are iid random variables, that they undergo inception in the site with the same initial size $N_0$, and that they grow exponentially with the same constant pre- and post-surgery net rates $\gamma_0$ and $\gamma_1$, respectively. Next, we employed the assumption that metastases are shed according to the Poisson process with rate r(t). Due to arbitrariness of the rate function r this essentially reduces to stochastic independence of the numbers of metastases shed over non-overlapping time intervals and the condition that the probability that multiple metastases were shed over an interval of small length h is o(h), see [23, pp. 281-284]. Finally, we postulated certain qualitative properties (specifically, continuity and positivity) of functions r and f, see Section 3. All these mathematical assumptions are what Henri Poincaré called *neutral hypotheses* [34]: they facilitate conceptualization and mathematical formalization of the phenomenon of interest without unduly restricting its scope.

Biomedical assumptions of our model are minor. One of them, that secondary metastasis is negligible, is based on the existing evidence [24]. Although mathematically possible, incorporation of secondary metastasis into the model would be impractical, for the generation number of a metastasis is largely unobservable. Another assumption posits that resection of primary tumor affects metastases through the rate of their growth rather than through a change in their latency times. This assumption is based on a well-known fact that quiescent cancer cells and dormant micro-tumors are refractory to various environmental signals and therapeutic interventions [27]. Interestingly, in the likelihood maximizing limiting case of zero metastatic growth prior to primary tumor excision, the phases of metastatic latency and pre-surgery growth become indistinguishable. Biologically, this amounts to the extension of metastatic latency until the time of surgery.

Given that the mechanisms by which primary tumor suppresses metastatic growth remain largely hypothetical [11, 28], a question arises as to how bearers of this trait were selected for in the course of evolution. In most cases primary tumor is not lethal *per se* while metastases are. Therefore, any spontaneously emerging mechanism of metastasis suppression by primary tumor would confer on the bearers of this trait survival and reproductive advantage over non-bearers while also allowing the former to pass it on to their offspring. As a result, the forces of natural selection would fix such mechanisms making them dominant across the relevant population.